\documentclass[journal]{IEEEtran}
\usepackage{}

\ifCLASSINFOpdf
\else
\fi
\usepackage{array}

\usepackage[cmex10]{amsmath}
\usepackage{bm}
\usepackage{enumerate}
\usepackage{amsfonts}
\usepackage{graphicx}
\usepackage{cite}
\usepackage{booktabs}
\usepackage{multirow}
\usepackage[linewidth=0.5pt]{mdframed}
\usepackage{threeparttable}
\usepackage{fixltx2e}
\usepackage{color}
\usepackage{xcolor}
\usepackage{subeqnarray}
\usepackage{cases}
\usepackage{makecell}
\usepackage{algorithmic}
\usepackage{algorithm}
\usepackage{bbold}
\ifCLASSOPTIONcompsoc
\usepackage[caption=false,font=normalsize,labelfon
t=sf,textfont=sf]{subfig}
\else
\usepackage[caption=false,font=footnotesize]{subfig}

\fi

\begin{document}
\bibliographystyle{IEEEtran}

\title{Fast Power System Cascading Failure Path Searching with High Wind Power Penetration}
\author{Yuxiao~Liu,~\IEEEmembership{Student Member,~IEEE,}
Yi~Wang,~\IEEEmembership{Member,~IEEE,}
Pei~Yong,~\IEEEmembership{Student Member,~IEEE,}
Ning~Zhang,~\IEEEmembership{Senior Member,~IEEE,}
Chongqing~Kang,~\IEEEmembership{Fellow,~IEEE,}
and~Dan~Lu

\thanks{ 
This work was supported in part by the National Key Research and Development Program of China under Grant 2016YFB0900100, the Major Smart Grid Joint Project of Natural Science Foundation of China and State Grid (No. U1766212), and Technical Project
of the State Grid: Theoretical and Empirical Research of the Key Technology for the Whole Process Management of Power Grid Operation Risk Based on Multi Source Data Mining. (\emph{Corresponding authors: Ning Zhang and Chongqing Kang})

Yuxiao Liu, Pei Yong, Ning Zhang and Chongqing Kang are with the Department of Electrical Engineering, Tsinghua University, Beijing 100084, China. (liuyuxiao16@mails.tsinghua.edu.cn; yp18@mails.tsinghua.edu.cn; ningzhang@tsinghua.edu.cn; cqkang@tsinghua.edu.cn)

Yi Wang is with the Power System Laboratory, ETH Zurich, 8092 Zurich, Switzerland. (yiwang@eeh.ee.ethz.ch)

Dan Lu is with the State Grid Henan Economic Research Institute, Zhengzhou 450052, China.
}
}
\maketitle


\begin{abstract}
Cascading failures have become a severe threat to interconnected modern power systems.
The ultrahigh complexity of the interconnected networks is the main challenge toward the understanding and management of cascading failures.
In addition, high penetration of wind power integration introduces large uncertainties and further complicates the problem into a massive scenario simulation problem.
This paper proposes a framework that enables a fast cascading path searching under high penetration of wind power.
In addition, we ease the computational burden by formulating the cascading path searching problem into a Markov chain searching problem and further use a dictionary-based technique to accelerate the calculations.
In detail, we first generate massive wind generation and load scenarios.
Then, we utilize the Markov search strategy to decouple the problem into a large number of DC power flow (DCPF) and DC optimal power flow (DCOPF) problems.
The major time-consuming part, the DCOPF and the DCPF problems, is accelerated by the dynamic construction of a line status dictionary (LSD).
The information in the LSD can significantly ease the computation burden of the following DCPF and DCOPF problems.
The proposed method is proven to be effective by a case study of the IEEE RTS-79 test system and an empirical study of China's Henan Province power system.
\end{abstract}

%
\IEEEpeerreviewmaketitle
\section{Introduction}
\begin{figure*}[hbt!]
	\centering
		\includegraphics[width=6.5in]{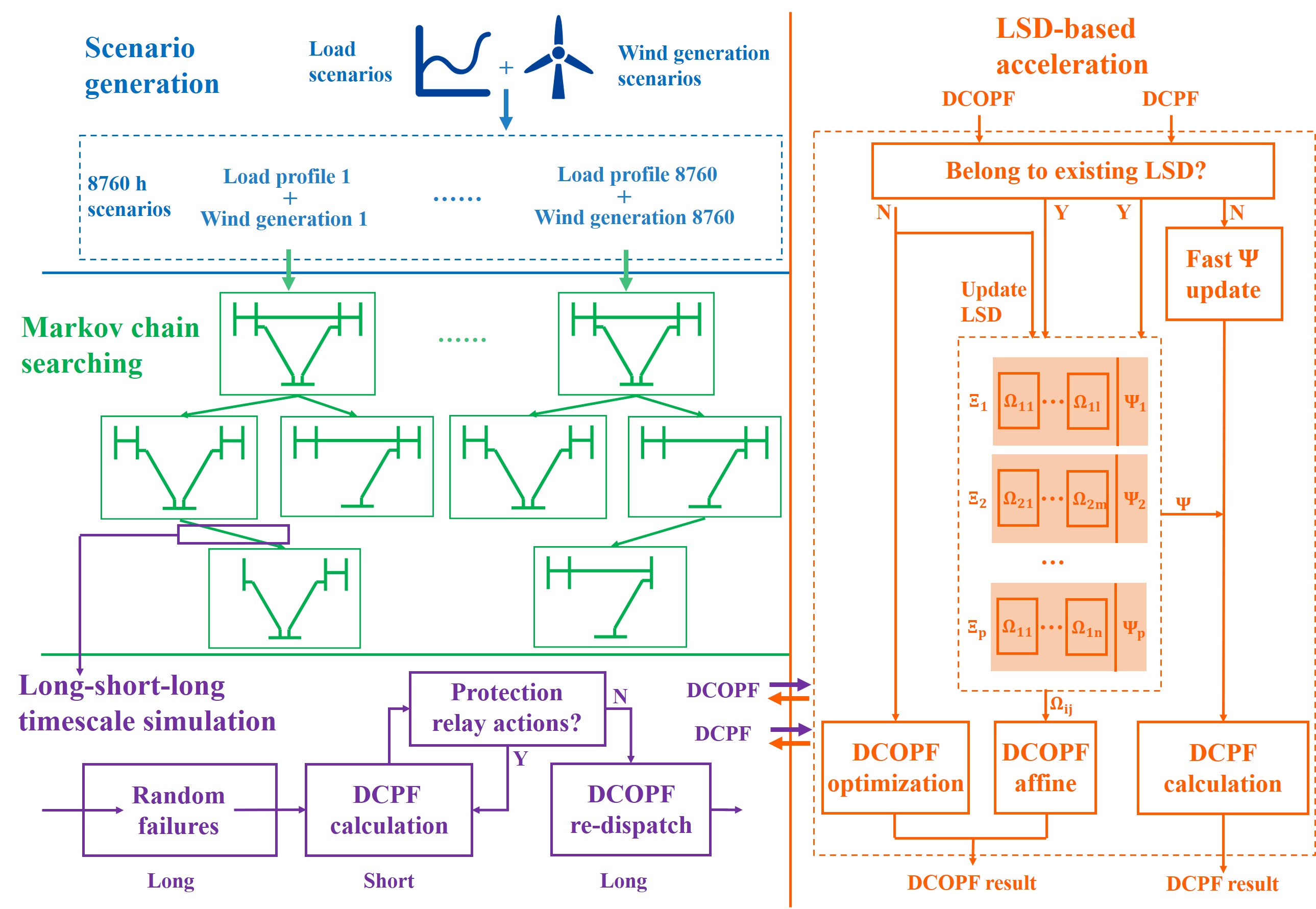}
\caption{The framework of the proposed method.}
	\label{fig1_framework}
\end{figure*}
\IEEEPARstart{A} cascading failure is a process in which the interconnected elements fail to work one after another dependently.
With the expansion of modern power systems, cascading failures can lead to severe damage to power systems and society \cite{hines2009large}.
A comprehensive and accurate searching of cascading failures can effectively help to evaluate the reliability and decrease the damage of failures in power systems, real-time operation, operational scheduling, and long-term planning processes \cite{vaiman2012risk,zlotecka2018characteristics}.

The paths of the power systems cascading failures are heavily related to the operation conditions, e.g., the load profiles and the wind power generation. Even the same initial failures may incur different cascading paths under different operation conditions.
Because of the increasing integration and high uncertainty of wind power generation, the patterns of power system operation conditions show great diversities \cite{wang2017efficient}.
As a result, the potential cascading failures become more uncertain with more unexpected consequences, making the searching method with the high penetration of wind generation increasingly essential\cite{athari2018impacts,wang2019reliability}.	

To consider the uncertainties and diversities of loads and wind generation, a large number of scenarios should be generated for a comprehensive simulation.
Each scenario requires a simulation of cascading failures.
Because of the complexity of power systems, even one scenario simulation of cascading failures has significant computational costs.
The cascading path searching incurs the problem of the ``curse of computational dimensionality" \cite{Billinton1994Reliability}; thus, the well-known Monte Carlo simulation method takes a long time to converge.
Hence, the computational burden when considering different load profiles and wind generation characteristics is unacceptable.

Some methods have been proposed to address the computational burden of cascading failure analysis.
One approach is to reduce the Monte Carlo samples by improving the sampling strategy so that more samples are simulated near the rare severe failures.
Such techniques include the splitting method \cite{wang2015efficient,kim2013splitting} and the importance sampling method \cite{henneaux2014improving, chen2013composite}.
Another approach formulates the cascading failure analysis as a Markov chain searching problem \cite{guo2018toward}.
The Markov chain searching problem can significantly accelerate the cascading failure simulation compared with the results of traditional sampling methods.
Because the Markov chain search turns the continuous and complex sampling into a discrete state space transition problem \cite{iyer2009markovian}, it avoids repeated calculations. In addition, the algorithm to support a high-throughput computing environment is another method to accelerate simulation \cite{anderson2017high}.

The influences of wind power uncertainties on cascading failures have recently been evaluated.
Anderson \textsl{et al.} showed that the increased integration and the uncertainty of wind power generation could trigger severe cascading failures \cite{athari2018impacts}.
They modeled the uncertainty of wind power generation by adopting a certain probability of line tripping by protection relays \cite{athari2019enhanced}.
Xu \textsl{et al.} carried out the cascading failure simulation using a multiscenario-based method and uncertainty modeling \cite{xu2018blackout}.
Sun \textsl{et al.} introduced a complex network theory for cascading failure analysis with wind integration \cite{sun2014cascading}.
The above works \cite{athari2018impacts,athari2019enhanced,xu2018blackout,sun2014cascading} model the wind power uncertainty in the day-ahead timescale.
That is, the wind forecasting results differ from the actual wind generation results.
Hence, the uncertainty can be modeled as the forecasting results plus a small zero mean signal \cite{athari2016modeling}.
In long-term timescales, e.g., one month or more, the accuracy of wind power generation prediction is not reliable; thus, the uncertainty cannot be modeled with the help of forecasting results.
The long-term timescale analysis of cascading failures with wind power integration is essential for operators to implement more strategies in advance \cite{vaiman2012risk}, but it has seldom been analyzed in previous research.

Cascading failure path searching with high wind power penetration benefit the power system by figuring out the key components mostly affect the security of the power system under wind uncertainty.
Different strategies could be applied when figuring out the key components (e.g. the branches or the generators) and the key scenarios (e.g. the bus power injection and branch flow scenarios).
In the short-term operation timescale, engineers can change the generations of some generators to avoid some key scenarios~\cite{babalola2018real,yao2018management}, or they can strengthen the inspection, and optimize the power system repairment resources of some key components~\cite{vaiman2012risk}.
In the long-term planning timescale, more strategies are available, such as building new transmission lines and new generation units~\cite{karimi2016considering}.

This paper addresses the problem of cascading failure searching with high penetration of wind power generation, considering the modeling of both the protection relays and the re-dispatch process.
We propose an acceleration approach that makes the computation of cascading failures searching acceptable in terms of the real-time operation timescale, the operation planning timescale, and even the long-term planning timescale (up to 8760 hours).
The proposed acceleration approach is based on the Markov chain searching approach and further constructs the line status dictionary (LSD) to simplify the computations.

This paper makes the following contributions:
\begin{enumerate}[1)]
\item Designing a framework to search power system cascading failure paths with high penetration of wind power generation, considering the potential failures under a sufficient number of wind generation and load consumption scenarios.
\item Proposing an LSD-based acceleration method that can significantly ease the computational burden of the massive scenario cascading simulation problem.
\item Conducting comprehensive case studies on the IEEE RTS-79 test system and the power system of China's Henan Province, to analyze the influence of high penetration of wind power on the power system cascading failures.
\end{enumerate}

The remainder of this paper is organized as follows.
Section \ref{section3_methodology} provides the detailed methodology of the searching of cascading failures, including scenario generation, Markov chain searching, cascading simulation, and LSD-based acceleration techniques.
Case studies are carried out in Section \ref{section4_casestudies} using the IEEE RTS-79 test system and the real-world power system of Henan Province.
Finally, Section \ref{section5_conclusions} draws the conclusions.


\section{Methodology of Cascading Failures Searching}
\label{section3_methodology}
The framework of the proposed cascading failures searching method is shown in \figurename\ref{fig1_framework}, which consists of four main processes:
1) The first process generates a sufficient number of wind and load scenarios (which will be introduced in Subsection \ref{Scenarios_Generation}).
2) Then follows the Markov chain search that obtains different states and possibilities in the cascading process (which will be introduced in Subsection \ref{Markov_Chain_Searching}).
3) Each step of the Markov chain search includes a long-short-long timescale simulation process that simulates the random failures (long), the protection relays (short), and the re-dispatch (long) process (which will be introduced in Subsection \ref{section_longshortlong}).
4) The main computation burden of the long-short-long timescale simulation is accelerated by the proposed LSD-based acceleration techniques (which will be introduced in Subsection \ref{section_LSD}).

\subsection{Scenario Generation}
\label{Scenarios_Generation}
To consider different operation conditions, we conduct scenario generation considering the sampling of load profiles and wind power generation.
The load profiles are simulated from historical load profiles and forecasted peak load level, which is modeled as uncontrollable in this work.
The proposed method could also consider controllable load by modeling the demand response as a virtual power generation unit \cite{huang2019demand}.
We generate the wind power generation scenarios using the wind power operation simulation module proposed in \cite{zhang2013planning}.
The scenarios are generated from the stochastic differential equations considering the wind speed probabilistic distribution, the spatial and temporal correlations, and the seasonal rhythms.
Afterwards, the wind generation scenarios, along with the load scenarios, are the input of the Markov chain searching process.
The scenarios, although generated from historical statistics, may not fully cover all the possible cases that lead to cascading failures.
One could apply different scenarios generation models to further observe possible cascading paths, which is out of the scope of this paper.

\subsection{Markov Chain Searching}
\label{Markov_Chain_Searching}
The purpose of the Markov chain searching is to obtain the states of the cascading process and the possibility of reaching the states.
The state here is defined as the combination of operational conditions (in operation or failure) of generators and lines.
We obtain each state and the corresponding possibility by sampling and simulation under different kinds of failures, which will be explained in detail in Section \ref{section_longshortlong}.

As shown in \figurename\ref{fig1_framework}, each node of the Markov chain denotes a single state, the arrows connecting different nodes denote long-short-long timescale simulations (which will be described in Subsection \ref{section_longshortlong}), and each combination of the connected nodes and arrows denotes a cascading path.
Each arrow has a certain possibility, which is calculated by the generator failures or line failure possibilities.
The possibility of reaching a state $n$ from a certain path $p$ equals the production of the arrows reaching the state $n$:
\begin{equation}\label{eq6_markov}
  p_{p,n}^{path}=p_p\left({{s}_{1}}\right)p_p\left({{s}_{2}}|{{s}_{1}}\right)...p_p\left({{s}_{n}}|{{s}_{n-1}}\right)
\end{equation}
where $p_p\left({{s}_{n}}|{{s}_{n-1}}\right)$ denotes the possibility of the arrow points from states ${s}_{n-1}$ to ${s}_{n}$.
Considering the fact that all the components may fail independently at any time, so that the cascading path with arbitrary independent failure combinations are theoretically possible.
The case that too many components fail randomly has ultrasmall possibilities and is out of concern in practice.
Hence, we set a threshold possibility $\varepsilon$ and consider only the paths that are higher than the threshold: $p_{i,n}^{path}\ge \varepsilon$.
This method can consider all the possible paths (including the dependent failures and independent failures) with the probability larger than threshold $\varepsilon$.
This threshold can be adjusted flexibly to balance the computation burden and the security level concerned in the searching process.
In addition, we consider the $m$ most severe paths with the most load shedding in each single scenario, with the probability larger than $\varepsilon$.
The cascading management plans are thereby made considering the most severe paths.

Note that compared with the sampling methods that randomly sample massive cascading cases, the Markov chain search can emerge the simulation of all the same cascading cases into the calculation of only one path, with a certain possibility of that path.
Thus, the Markov chain searching method can avoid repeatedly calculating the same cases and is more computationally efficient.

\subsection{Long-short-long Timescale Simulation}
\label{section_longshortlong}

The long-short-long timescale simulation is the basic component in the Markov chain searching process, which is composed of 1) long timescale failures triggered by random failures of generators and lines; 2) short timescale failures triggered by protection relays; and 3) long timescale failures representing the re-dispatch process\cite{yao2017risk}.
Each short timescale failure lasts only several seconds, and each long timescale failure lasts around 10 minutes.
Note that the wind generation considered in this work is substation level, where for simplicity, the generation fluctuation is assumed as unchanged during the cascading process.
Our acceleration technique is also applicable when changing the wind generation scenarios during the cascading process.

\subsubsection{Random failures}
\noindent

The first long timescale simulation considers random failures of generators and lines.
This step enumerates all the possible single component failure with the probability larger than the preset threshold $\varepsilon$.
The random failures of generators follow a certain probability that can be modeled using the statistics of history failures.
Note that we only model the conventional generators in this way.
The failures of wind power units are considered in the model\cite{zhang2013planning} of wind scenarios generation process.
The random failures of lines are modeled by both the statistics of history failures and the branch flow level of the transmission lines.
In detail, the failure probability of line $k$ is as follows:
\begin{equation}
  \begin{aligned}\label{eq1_line_failure}
    {{p}_{k}}=\left\{ \begin{matrix}
      p_{k}^{b} & \left| {{L}_{k}} \right|\le L_{k}^{b}  \\
      \left(1-p_{k}^{b}\right)\times {{L}_{k}}/\beta_{k} L_{k}^{b}+p_{k}^{b} & L_{k}^{b}<\left| {{L}_{k}} \right|\le \beta_{k} L_{k}^{b}  \\
      1 & \left| {{L}_{k}} \right|>\beta_{k} L_{k}^{b}  \\
   \end{matrix} \right.     
  \end{aligned}
\end{equation}
where $p_{k}^{b}$ denotes the basic failure probability of line $k$, ${L}_{k}/L_{k}^{b}$ denote the actual$/$basic branch flow of line $k$, and $\beta_{k}$ denotes the protection relay threshold of line $k$.
Equation \eqref{eq1_line_failure} states that ${p}_{k}$ becomes larger when the branch flow ${L}_{k}$ exceeds the basic value $L_{k}^{b}$, until ${L}_{k}$ reaches the value that triggers the protection relay.
Note that the increasing integration of wind generation significantly changes the patterns of branch flows and thus changes the failure probability of lines.

\subsubsection{Protection relays}
\noindent

The short timescale simulation realizes the process triggered by protection relay after the random failures.
The protection relay  triggers the circuit breakers and removes the line from services after detecting the branch flow violations.
To model this process, we conduct the DCPF, disconnect the lines that trigger the circuit breaker, and then conduct the DCPF again until no circuit breaker is operating.
The DCPF equations are formulated by the generation shifted distribution factor (GSDF):
\begin{equation}\label{eq2_DCPF}
  \bm{L}=\mathbf{\Psi}\bm{P}
\end{equation}
where $\bm{L}$, $\bm{P}$, and $\mathbf{\Psi}$ denote the branch flow vector, the bus active injection vector that includes both generators and loads, and the GSDF matrix, respectively.
The GSDF is a $K \times N$ matrix, where $K$ and $N$ are the number of lines and buses, respectively.
This matrix can be constructed by first calculating the $K \times (N-1)$ matrix $\mathbf{\tilde{\Psi }}$:
\begin{equation}\label{eq3_GSDF}
  \mathbf{\tilde{\Psi }}=\mathbf{\tilde{B}}_{\mathbf{f}}{\mathbf{\tilde{B}}}^{\mathbf{-1}}
\end{equation}
then inserting the zero column at the column corresponding to the reference bus (which also can be set as other buses).
In \eqref{eq3_GSDF}, $\mathbf{\tilde{B}}_{\mathbf{f}}$ and $\mathbf{\tilde{B}}$ denote the matrix $\mathbf{B}_{\mathbf{f}}$ and $\mathbf{B}$ after removing the column and the row of the reference bus, respectively.
$\mathbf{B}_{\mathbf{f}}$ and $\mathbf{B}$ are the branch matrix and bus matrix constructed by the reciprocal of the line reactance in the DCPF theory, respectively; see \cite{stott2009dc} for details.

In the simulation process, the GSDF matrix $\mathbf{\Psi }$ is updated with the change in the system topologies.
However, massive calculations of matrix inversion in \eqref{eq3_GSDF} cost substantial time.
We implement a fast update method of the GSDF to shorten the time consumption.
Because each time $\mathbf{\tilde{B}}$  has only small changes compared to the initial $\mathbf{\tilde{B}}$ without line failures, the update of $\mathbf{\Psi }$ can be simplified by the matrix inversion lemma\cite{woodbury1950inverting}.
Define the $K \times 1$ binary line state vector $\mathbf{S_L}$ with 0-down and 1-operational, and define the $N \times L$ node-branch incident matrix $\mathbf{E}$.
The $k$th column of $\mathbf{E}$ is:
\begin{equation}
    \mathbf{E_k} = {{\left[ \underset{1}{\mathop{0}}\,\text{ }...\text{ }\underset{k_i}{\mathop{\text{1}}}\,\text{ }...\text{ }\underset{k_j}{\mathop{\text{-1}}}\,\text{ }...\text{ }\underset{N}{\mathop{\text{0}}}\, \right]}^{T}}
\end{equation}
where $k_i$ and $k_j$ is the start and end bus of line $k$, respectively.
Given a state $\mathbf{S_L}$, formulate set $K$:$\left\{ k\in K|{{\mathbf{S}}_{\mathbf{L}}}(k)=0 \right\}$.
Collect all the $\mathbf{E_k}$ that exist in the set $K$ to obtain $\mathbf{M}$:
\begin{equation}
    \mathbf{M} = {{\left[\text{ }...\text{ }{\mathop{\mathbf{E_k}}}\,\text{ }...\, \right]}}\text{ }k \in K
\end{equation}
Then, remove the row corresponding to the reference bus and formulate $\mathbf{\tilde{M}}$.
Subsequently, the update of $\mathbf{\tilde{\Psi }}$ can be simplified as:
\begin{subequations}\label{eq5_matrix_inversion}
  \begin{equation}\label{eq5a_matrix_inversion}
    \begin{aligned}
      {{{\mathbf{\tilde{\Psi }}}}^{\text{new}}}&=\mathbf{\tilde{B }}_{\mathbf{f}}^{\text{new}}{{\left({{{\mathbf{\tilde{B}}}}^{\text{new}}}\right)}^{-1}} \\ 
      &=\mathbf{\tilde{B }}_{\mathbf{f}}^{\text{new}}{{\left(\mathbf{\tilde{B}}-\mathbf{\tilde{M}}\mathbf{b}_\mathbf{k}^{-1}{{{\mathbf{\tilde{M}}}}^{\mathbf{T}}}\right)}^{-1}} \\ 
      &=\mathbf{\tilde{B }}_{\mathbf{f}}^{\text{new}}\left(\mathbf{\tilde{B}}-{{{\mathbf{\tilde{B}}}}^{-1}}\mathbf{\tilde{M}c}{{{\mathbf{\tilde{M}}}}^{\mathbf{T}}}{{{\mathbf{\tilde{B}}}}^{-1}}\right) \\ 
      &=\mathbf{\tilde{\tilde{\Psi }}}-\mathbf{\tilde{B }}_{\mathbf{f}}^{\text{new}}{{{\mathbf{\tilde{B}}}}^{-1}}\mathbf{\tilde{M}c}{{{\mathbf{\tilde{M}}}}^{\mathbf{T}}}{{{\mathbf{\tilde{B}}}}^{-1}} \\ 
    \end{aligned}
  \end{equation}
  \begin{equation}\label{eq5b_matrix_inversion}
    \text{where }\mathbf{c}\text{=}{{\left[ -\mathbf{b}_\mathbf{k}^{-1}+{{{\mathbf{\tilde{M}}}}^{\mathbf{T}}}{{{\mathbf{\tilde{B}}}}^{-1}}\mathbf{\tilde{M}} \right]}^{-1}}
  \end{equation}
\end{subequations}
where the superscript ``new'' denotes the updated matrix, $\mathbf{\tilde{\tilde{\Psi }}}$ denotes the original $\mathbf{\tilde{\Psi }}$ matrix with the row corresponding to line $k$ removed, and ${b}_{k}$ denotes the reciprocal of the line $k$ reactance.
When updating ${{\mathbf{\tilde{\Psi }}}}$ with $l$ line failures, ${b}_{k}$ is an $l \times l$ diagonal matrix, $\mathbf{M}$ is a matrix with $l$ columns, and the other derivations are the same.
We can conclude that the inverse calculation of the $N \times N$ matrix $\mathbf{\tilde{B}}$ in \eqref{eq3_GSDF} turns into the inverse calculation of the $l \times l$ diagonal matrix and several matrix multiplications, which significantly improves the computational complexity.

\subsubsection{Re-dispatch}
\noindent

The second long-timescale simulation is the re-dispatch of the power flow after the random failures and the action of protection relays.
This process is used to determine how much load shedding is needed to maintain the reliability of the system.
Each re-dispatch process can be formulated by a DCOPF model:
\begin{subequations}\label{eq5_dcopf}
\begin{equation}\label{eq5a_dcopf} 
  \underset{\bm{P},\bm{dD}}{\mathop{\min }}\,\mathbf{c}_{\mathbf{p}}^{\mathbf{T}}\bm{P}+\mathbf{c}_{\mathbf{d}}^{\mathbf{T}}\bm{dD}
\end{equation}
\begin{equation}\label{eq5b_dcopf}
  {{\mathbf{1}}^{\mathbf{T}}}\bm{P}=0
\end{equation}
\begin{equation}\label{eq5c_dcopf}
  -{{\mathbf{L}}^{\mathbf{b}}}\text{diag}\left(\mathbf{S_L}\right)\le \mathbf{\Psi}\bm{P}\le {{\mathbf{L}}^{\mathbf{b}}}\text{diag}\left(\mathbf{S_L}\right)
\end{equation}
\begin{equation}\label{eq5d_dcopf}
  \bm{0}\le \bm{dD}\le \mathbf{D}
\end{equation}
\begin{equation}\label{eq5e_dcopf}
  \bm{P}-\bm{dD}\le \mathbf{C}\text{diag}\left( {{\mathbf{G}}_{\mathbf{max}}} \right){{\mathbf{S}}_{\mathbf{G}}}-\mathbf{D}
\end{equation}
\end{subequations}
where $\bm{P}$ and $\bm{dD}$ are unknown variables that denote the bus active injection vector and the load-shedding vector, respectively;
 $\mathbf{D}$ and ${\mathbf{S}}_{\mathbf{G}}$ are vectors that denote the loads and the binary states of generators with 0-down and 1-operational, respectively;
 $\mathbf{c}_{\mathbf{P}}/\mathbf{c}_{\mathbf{d}}$, ${\mathbf{L}}^{\mathbf{b}}$, and $\mathbf{G}_\mathbf{max}$ are vectors that denote the costs of generation$/$load-shedding, the long-term maximum branch flows, and the maximum generator output, respectively;
and $\mathbf{C}$ is the matrix that denotes the system's initial generation connection matrix.

In \eqref{eq5_dcopf}, the objective function \eqref{eq5a_dcopf} minimizes the summation costs of generation and load shedding.
\eqref{eq5b_dcopf} is the active power lossless constraint of the DCPF equations.
\eqref{eq5c_dcopf} and \eqref{eq5d_dcopf} denote the branch flow constraint and the load-shedding constraint, respectively.
Constraint \eqref{eq5e_dcopf} denotes the nodal power balance, indicating that the power generation of a bus should be below the maximum generator output connected to the bus.
Note that the $\mathbf{G}_\mathbf{max}$ in constraint \eqref{eq5e_dcopf} includes the wind power generation.
In each scenario, the wind power generation is a constant, and thus $\mathbf{G}_\mathbf{max}$ is a constant.
In different scenarios, however, the $\mathbf{G}_\mathbf{max}$ changes with the change in wind power generation.
Hence, a sufficient number of scenarios is required to model the wind power uncertainty, which incurs the problem of a high computational burden.

\subsection{LSD-based Acceleration Techniques}
\label{section_LSD}

The purpose of the LSD-based acceleration techniques is to ease the computational burden of a large number of long-short-long timescale simulations since massive scenarios and massive cascading paths in the Markov chain searching need to be calculated.
The main computational burden of long-short-long timescale simulation comes from the DCOPF calculations and the DCPF calculations.
In this paper, we conduct the LSD-based technique to accelerate the calculation.
On the one hand, we use the multi-parametric linear programming (MPLP) method based on the LSD to accelerate the calculation of the DCOPF.
On the other hand, we strategically store the GSDF matrix in the LSD to accelerate the DCPF calculation.

\subsubsection{The construction of the LSD}
\noindent

The LSD-based technique is inspired by the fact that both the formulations of the DCOPF and DCPF calculations are highly related to the line state of the power system.
Calculations can be simplified by the existing calculation results of the same line state.
Hence, we build the LSD and store the calculation results by line states.

As shown in \figurename\ref{fig1_framework}, for each line state ${\mathbf{\Xi }}_{\mathbf{i}}$, the LSD consists of a set of critical regions $\mathbf{\Omega }_\mathbf{ij}$ and one GSDF matrix $\mathbf{\Psi }_\mathbf{i}$:
\begin{equation}\label{eq11_LSD}
  \left\{ {{\mathbf{\Xi }}_{\mathbf{i}}}|{{\mathbf{\Omega }}_{\mathbf{ij}}},{{\mathbf{\Psi }}_{\mathbf{i}}}\text{ with }\mathbf{j}=\mathbf{1}\text{...}\mathbf{J} \right\}
\end{equation}
where $\mathbf{\Omega }_\mathbf{ij}$ represents the $\mathbf{j}$th critical region in the $\mathbf{i}$th line state.
1) The construction of critical regions $\mathbf{\Omega }_\mathbf{ij}$ is used to accelerate the DCOPF calculation, which will be elaborated in the following subsection.
2) The construction of a GSDF matrix ${\mathbf{\Psi }}_{\mathbf{i}}$ is used to accelerate the DCPF calculation.

In the cascading searching process, we dynamically construct the LSD in its entirety.
In each DCOPF calculation, we update the results in \eqref{eq11_LSD} if they are not in the LSD.
Otherwise, we use the result in the LSD to accelerate the calculation.


\subsubsection{The MPLP method}
\noindent

The MPLP method is used to accelerate DCOPF calculations.
MPLP can simplify a linear programming (LP) optimization problem into an affine calculation problem.
In detail, an LP problem can be formulated as:
\begin{subequations}\label{eq7_mplp_lp}
\begin{equation}\label{eq7a_mplp_lp}
  \underset{\bm{x}}{\mathop{\min }}\,\mathbf{c}\bm{x}
\end{equation}
\begin{equation}\label{eq7b_mplp_lp}
  \mathbf{A}\bm{x}\le \mathbf{b}+\mathbf{\Delta b}=\mathbf{b}+\mathbf{F}\bm{\varphi } 
\end{equation}
\end{subequations}
where $\mathbf{\Delta b}$ denotes the changing part of $\mathbf{b}$ among different optimization problems.
The changing part is represented by a constant matrix $\mathbf{F}$ and a parameter vector $\bm{\varphi}$.
Thus, the optimal solution $\bm{x}^{*}(\bm{\varphi})$ is a function of $\bm{\varphi}$.
The constraints in \eqref{eq7b_mplp_lp} can be divided into active constraints with \eqref{eq8a_mplp_two_constraints} that reach the boundary and inactive constraints with \eqref{eq8b_mplp_two_constraints} that do not reach the boundary.
\begin{subequations}\label{eq8_mplp_two_constraints}
  \begin{equation}\label{eq8a_mplp_two_constraints}
    \mathbf{\tilde{A}}{{\bm{x}}^{\bm{*}}}(\bm{\varphi })=\mathbf{\tilde{b}}+\mathbf{\tilde{F}}\bm{\varphi }
  \end{equation}
  \begin{equation}\label{eq8b_mplp_two_constraints}
    \mathbf{\bar{A}}{{\bm{x}}^{\bm{*}}}(\bm{\varphi })<\mathbf{\bar{b}}+\mathbf{\bar{F}}\bm{\varphi }
  \end{equation}
\end{subequations}
The following inequality is satisfied if a new LP problem with $\bm{\varphi^{new}}$, $\mathbf{\tilde{F}^{new}}$ and $\mathbf{\bar{F}^{new}}$ has the same active and inactive constraints of \eqref{eq8a_mplp_two_constraints} and \eqref{eq8b_mplp_two_constraints}:
\begin{equation}\label{eq8.2_judge}
  \left(\mathbf{\bar{A}} \mathbf{\tilde{A}^{-1}} \mathbf{\tilde{F}^{new}}-\mathbf{\bar{F}^{new}} \right) \bm{\varphi ^{new}}<\mathbf{\bar{b}}-\mathbf{\bar{A}} \mathbf{\tilde{A}^{-1}} \mathbf{\tilde{b}}
\end{equation}
The optimization results are  decided by the active constraints only.
Thus, define problems that have the same active constraints are in the same critical region $\mathbf{\Omega }_\mathbf{ij}$:
\begin{equation}\label{eq11.5_LSD}
  \left\{ {{\mathbf{\Omega }}_{\mathbf{ij}}}|\mathbf{\tilde{A}}_{\mathbf{ij}}^{\mathbf{-1}},{{{\mathbf{\tilde{b}}}}_{\mathbf{ij}}} \right\}
\end{equation}
Under the theory of MPLP, the optimal solutions of optimization problems in the same critical region can be given by a simple affine calculation:
\begin{equation}\label{eq8.5_affine}
  {{\bm{x}}^{\mathbf{*}}}(\bm{\varphi^{new}})={{\mathbf{\tilde{A}}}^{-1}}\left( \mathbf{\tilde{b}}+\mathbf{\tilde{F}^{new}}\bm{\varphi ^{new}} \right)
\end{equation}
See \cite{borrelli2003geometric,ji2017probabilistic} for detailed demonstrations.
As a result, we replace the original optimization calculation (often using the interior-point method with a considerable number of iterations\cite{wright2005interior}) with the simple affine calculation for LP problems in the same critical region.

In each single line status, the GSDF matrix $\mathbf{\Psi}$ remains unchanged; thus, the DCOPF problem can be simplified by MPLP.
In detail, for DCOPF problems in the same element of the LSD set ${{\mathbf{\Xi }}_{\mathbf{i}}}$ and the same critical region set ${\mathbf{\Omega }}_{\mathbf{ij}}$, we obtain the $\mathbf{\tilde{A}}_{\mathbf{ij}}^{\mathbf{-1}}$ and ${{{\mathbf{\tilde{b}}}}_{\mathbf{ij}}}$ stored in the LSD and apply the affine calculation in \eqref{eq8.5_affine} to obtain the results of the DCOPF.
Otherwise, we use the interior-point method to calculate the result and update the LSD.

To accord with the formulation of \eqref{eq7b_mplp_lp}, we reformulate the inequality constraints in \eqref{eq5c_dcopf}-\eqref{eq5e_dcopf} as:
\begin{equation}\label{eq9_mplp_constraint}
  \!\left[ \begin{matrix}
    \mathbf{\Psi }  \\
    -\mathbf{\Psi }  \\
    \mathbf{I}  \\
    \mathbf{0}  \\
    \mathbf{0}  \\
 \end{matrix}\begin{matrix}
    \mathbf{0}  \\
    \mathbf{0}  \\
    -\mathbf{I}  \\
    \mathbf{I}  \\
    -\mathbf{I}  \\
 \end{matrix} \right]\!\!\!\left[ \begin{matrix}
    \bm{P}  \\
    \bm{dD}  \\
 \end{matrix} \right]\!\!\le\!\! \left[ \begin{matrix}
    {{\mathbf{L}}^{\mathbf{b}}}\text{diag}({{\mathbf{S}}_{\mathbf{L}}})  \\
    -{{\mathbf{L}}^{\mathbf{b}}}\text{diag}({{\mathbf{S}}_{\mathbf{L}}})  \\
    \mathbf{0}  \\
    \mathbf{0}  \\
    \mathbf{0}  \\
 \end{matrix} \right]\!\!+\!\!\left[ \begin{matrix}
    \mathbf{0}  \\
    \mathbf{0}  \\
    \mathbf{C}  \\
    \mathbf{0}  \\
    \mathbf{0}  \\
 \end{matrix}\begin{matrix}
    \mathbf{0}  \\
    \mathbf{0}  \\
    -\mathbf{I}  \\
    \mathbf{I}  \\
    \mathbf{0}  \\
 \end{matrix} \right]\!\!\!\left[ \begin{matrix}
    \text{diag}\left( {{\mathbf{G}}_{\mathbf{max}}} \right){{\mathbf{S}}_{\mathbf{G}}}  \\
    \mathbf{D}  \\
 \end{matrix} \right] 
\end{equation}
The inequality constraints in \eqref{eq9_mplp_constraint} are added into the active constraints \eqref{eq8a_mplp_two_constraints} or inactive constraints \eqref{eq8b_mplp_two_constraints}, depending on whether the boundary is reached.
The equality constraint \eqref{eq5a_dcopf} is directly added into the active constraint in \eqref{eq8a_mplp_two_constraints}.


\subsubsection{The Storage of the GSDF Matrix}
\noindent

Storing the GSDF matrix $\mathbf{\Psi}$ can further simplify the calculation of the DCPF and DCOPF.
The $\mathbf{\Psi}$ that is calculated in the existing DCOPF can be directly used in the DCPF and DCOPF calculations.
The update of $\mathbf{\Psi}$ in the DCPF calculations in \eqref{eq5_matrix_inversion} and the DCOPF calculations in \eqref{eq9_mplp_constraint} is thereby reduced.

\subsubsection{Implementing Other Simulations}
\noindent

The proposed acceleration techniques can be applied when some other protection simulations are considered.
In such cases, we conduct the same long-short-long timescale simulation framework consists of power flow and optimal power flow calculations.
The LSD-based acceleration is applicable as long as the optimal power flow problems are in convex formulation, so that we can implement the multi-parametric programming framework~\cite{dominguez2010recent}.
Recently, some convex power flow models are proposed with high accuracy~\cite{liu2018data,barzegar2019method}.
Such convex models can address the calculation of voltage magnitudes and reactive power, and thus enable the simulation of more protections triggered by voltage magnitudes or reactive power.
Therefore, we can further consider the following protections using the proposed acceleration techniques: line protections triggered by under voltage conditions, generator failures triggered by under or over voltage conditions, and generator protections triggered by reverse reactive power, etc.
Still, this work is not designed for the acceleration of dynamic simulations, such as the protection triggered by small signal instability.


\section{Case Studies}
\label{section4_casestudies}
In this section, the proposed method is tested on the IEEE RTS-79 test system and the power system of Henan Province.
The computation time comparison of different approaches and the influence of wind generation on cascading results are demonstrated in this section.
\subsection{IEEE RTS-79 test system}
\subsubsection{System Description}
\noindent

The IEEE RTS-79 system was proposed for risk-based analysis.
This system contains 24 buses, 38 lines, and 32 generators with a peak load of 2850 MW and a total generation capacity of 3405 MW.
Detailed information about the system can be found in \cite{subcommittee1979ieee}.

To simulate the case with a high share of wind power integration, 5 wind farms, Gen \#34-\#38, are connected to the grid.
They are connected to Bus \#1, \#2, \#18, \#21, and \#23, respectively, with a capacity of 340 MW each.
Correspondingly, we remove the thermal generator Gen \#2-3, Gen \#6-7, and Gen \#31 from Bus \#1, \#2, and \#23 with the capacity of 152 MW, 152 MW, and 155 MW, respectively.
Considering the geographic location of the buses, set the correlation coefficients of the wind speed of the wind farms by regions.
Bus \#1, \#2, and \#3 are in the same region, and Bus \#4 and \#5 are in the same region.
We set the correlation coefficients of the buses in the same region as 0.6 and the correlation coefficients of the buses in different regions as 0.2.

We set the minimum simulation threshold as $\varepsilon = 10^{-9}$.
A total of 8760 scenarios are generated to consider the load and wind generation uncertainties.
The number of cascading paths considered in each scenario is set as $m=3$.
Thus, we have $8760 \times 3=26280$ cascading paths after the searching.
\subsubsection{Time Consumption Analysis}
\noindent

The simulation of cascading failures was carried out using MATLAB 2016b, Cplex 12.4\cite{cplex}, and Matpower 6.0\cite{zimmerman2011matpower} on a standard PC with an Intel\normalsize{\small{\textcircled{\footnotesize{R}}}} $\text{Core}^\text{TM}$ i7-6700HQ CPU running at 2.60 GHz and 16.0 GB of RAM.
The time consumption of three cases is compared:
\begin{enumerate}[1)]
\item \textbf{c1:} The case without the construction of the LSD or the fast update of the GSDF.
\item \textbf{c2:} The case with the construction of the LSD, without the fast update of the GSDF.
\item \textbf{c3:} The case with the construction of the LSD and the fast update of the GSDF.
\end{enumerate}
\begin{table}[ht]
  \centering
  \caption{Time Consumptions of Different Cases}
  \label{table2_time_consumptions}
  \begin{tabular}{@{}ccccc@{}}
  \toprule
  \multicolumn{1}{l}{\begin{tabular}[c]{@{}l@{}}Time cons-\\umption (s)\end{tabular}} & \multicolumn{1}{l}{\begin{tabular}[c]{@{}l@{}}Random failure\\ sampling\end{tabular}} & \multicolumn{1}{l}{\begin{tabular}[c]{@{}l@{}}DCPF \\calculation\end{tabular}} & \multicolumn{1}{l}{\begin{tabular}[c]{@{}l@{}}DCOPF \\re-dispatch\end{tabular}} & \multicolumn{1}{l}{\begin{tabular}[c]{@{}l@{}}Total\\time\end{tabular}} \\ \midrule
  \textbf{c1} & 82 & 241 & 1425 & 1883\\
  \textbf{c2} & 74 & 124 & 94 & 351\\
  \textbf{c3} & 74 & 78 & 91 & 297\\ 
\bottomrule
  \end{tabular}
\end{table}

The results are depicted in Table \ref{table2_time_consumptions}. It can be concluded that the proposed method improves the computational efficiency from 1833 s to 297 s.
Compared with \textbf{c1}, \textbf{c2} demonstrates the effect of the construction of the LSD, which significantly reduces 93\% of the DCOPF re-dispatch calculation time and 49\% of the DCPF calculation time.
Compared with \textbf{c2}, \textbf{c3} demonstrates the effect of the GSDF fast update, with 37\% acceleration of the DCPF calculation.
Such improvements would enable a more efficient cascading failures searching under the same computation hardware in the process of both power system real-time operation and operation planning.

\subsubsection{The Influence of Wind Power Generation}
\noindent

\begin{figure*}[htb!]
	\centering
		\includegraphics[width=7.2in]{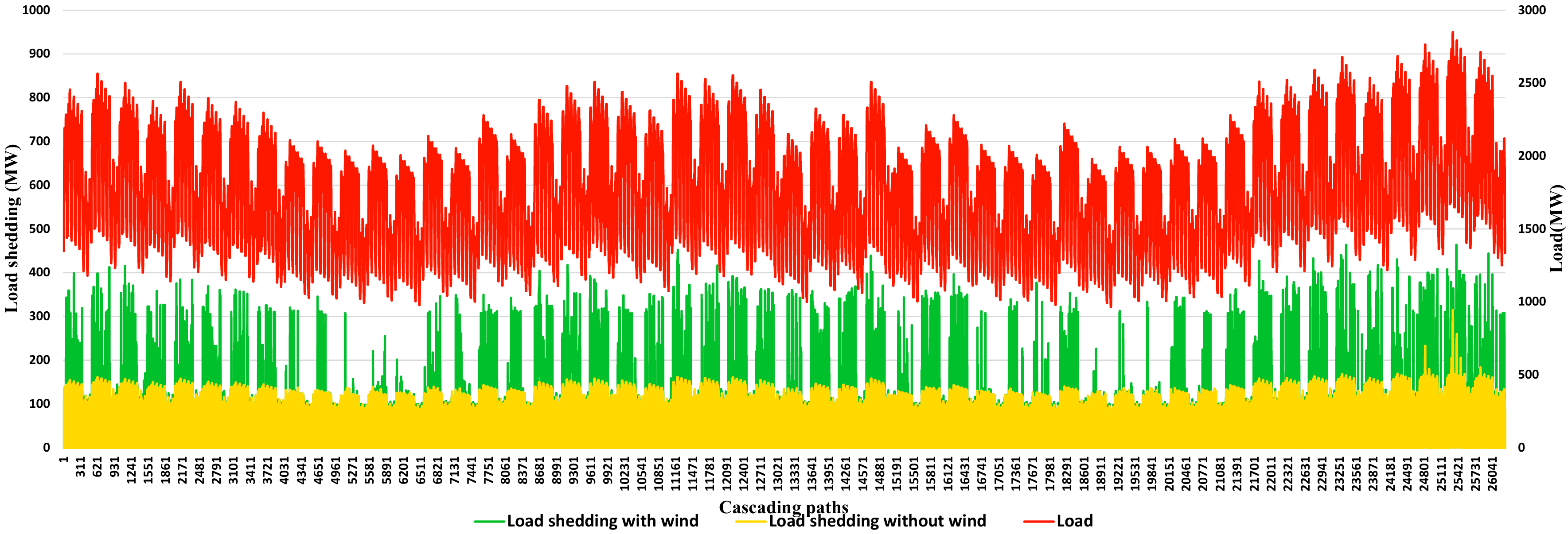}
\caption{The load shedding and the load in different cascading paths of the IEEE RTS-79 test system.}
	\label{fig2_8760hours}
\end{figure*}

We demonstrate the overall load shedding results and the cascading paths of all simulation scenarios and the detailed results of some typical cascading paths.

The overall load shedding results of all simulation scenarios are depicted in \figurename\ref{fig2_8760hours}.
The figure compares the results of load shedding with and without wind (scaled to the left vertical axis), and the total load of the system (scaled to the right vertical axis) in terms of all 26280 simulated cascading paths throughout the year.
From \figurename\ref{fig2_8760hours}, the maximum load sheddings of all cascading paths increase from 314MW to 464MW when considering wind power generation.
In addition, the amount of load shedding without wind is highly correlated with the total load of the system, and the largest load shedding occurs at the maximum load hour.
Thus, the system operators usually conduct the simulation only at the maximum load hour.
Regarding the result with wind integration, however, the load shedding is significantly larger than the case without wind integration.
More importantly, the amount of load shedding has far less correlation with the total load of the system.
This finding illustrates that the wind integration introduces great uncertainties into the results of the cascading failures.
Hence, conducting the cascading failures searching with a high share of wind integration is essential under a massive number of scenarios.

\begin{figure*}[hbt!]
	\centering
		\includegraphics[width=7.2in]{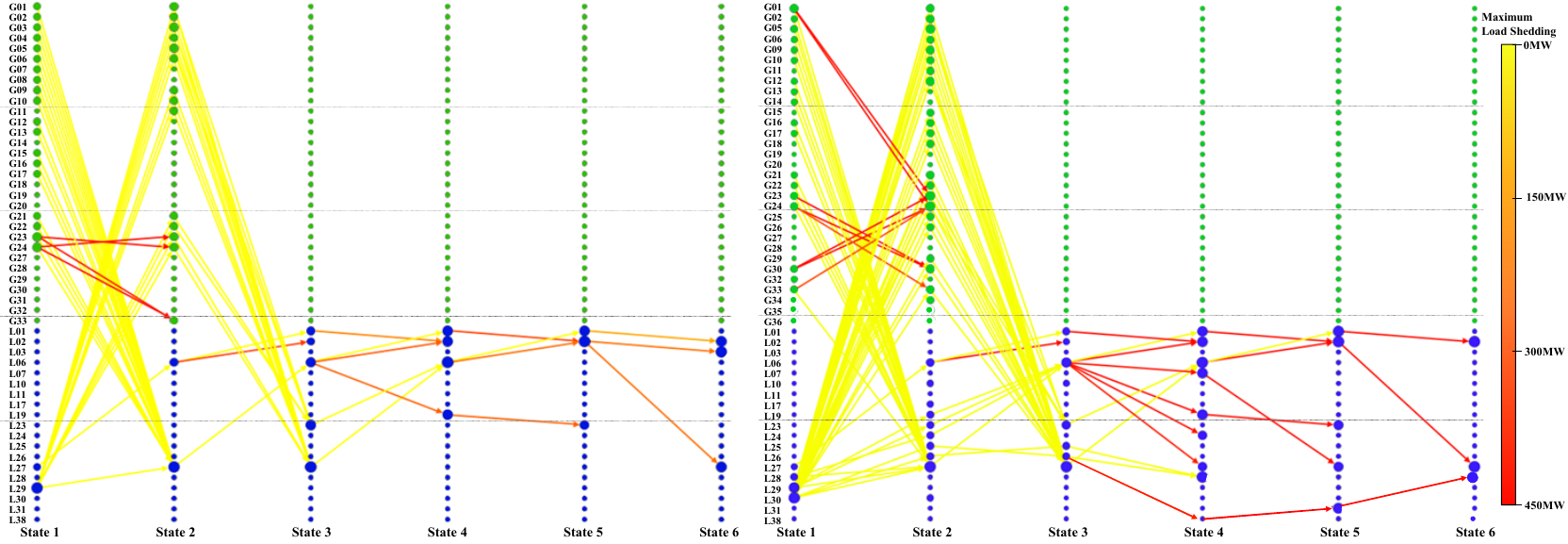}
\caption{The cascading paths of the IEEE RTS-79 system without (left) and with (right) wind integration.}
	\label{fig3_wholepicture}
\end{figure*}

The cascading paths of all simulation scenarios are shown in \figurename\ref{fig3_wholepicture}.
The case without wind integration is shown on the left, and the case with wind integration is shown on the right.
All simulated paths are depicted in the manner of nodes and arrows.
Each node represents an element failure in the IEEE RTS-79 system.
For example, G01 denotes the failure of Gen \#01, and L01 denotes the failure of Line \#01.
The nodes are distributed in columns, with each column representing a state in a cascading path.
It can be considered that each state of the cascading path has only one element failure.
Thus, a cascading path can be represented by arrows pointing from one node in one column to another node in another column.
For example, the cascading path ``L27$\rightarrow$L06$\rightarrow$L02'' denotes that the element failure happens in the order of Line \#27, Line \#06, and Line \#02.
The size of the nodes that each column represents is proportional to the degree (the number of arrows connected to the node) of the node.
In each column, the node with a larger degree has a larger size.
The color of the arrows represents the maximum load shedding from one state to another state.
The load shedding increases from yellow to red.
It should be noted that elements that never fail in the simulation are omitted in the figure to save space.

\figurename\ref{fig3_wholepicture} shows a whole evolution process among all elements and of all simulated paths.
We show that all 26280 cascading paths can be clearly demonstrated by only hundreds of arrows.
This approach gives the system operator a comprehensive visualization: what are the possible cascading paths, how are the cascading paths triggered, and what is the severity of each element failure in the paths.
It can be observed that the cascading paths that have more element failures end up with continuous line failures instead of generator failures.
In addition, the case that considers wind integration has more arrows, especially more red arrows.
New types of cascading paths occur with wind integration, such as ``L06$\rightarrow$L07'' and ``L38$\rightarrow$L31''.
Hence, the cascading paths become more complicated and more severe in the case that considers wind integration.

\begin{figure*}[ht!]
	\centering
\begin{minipage}[ht]{8.8cm}
		\centering
		\includegraphics[width=2.3in, height=3.5in]{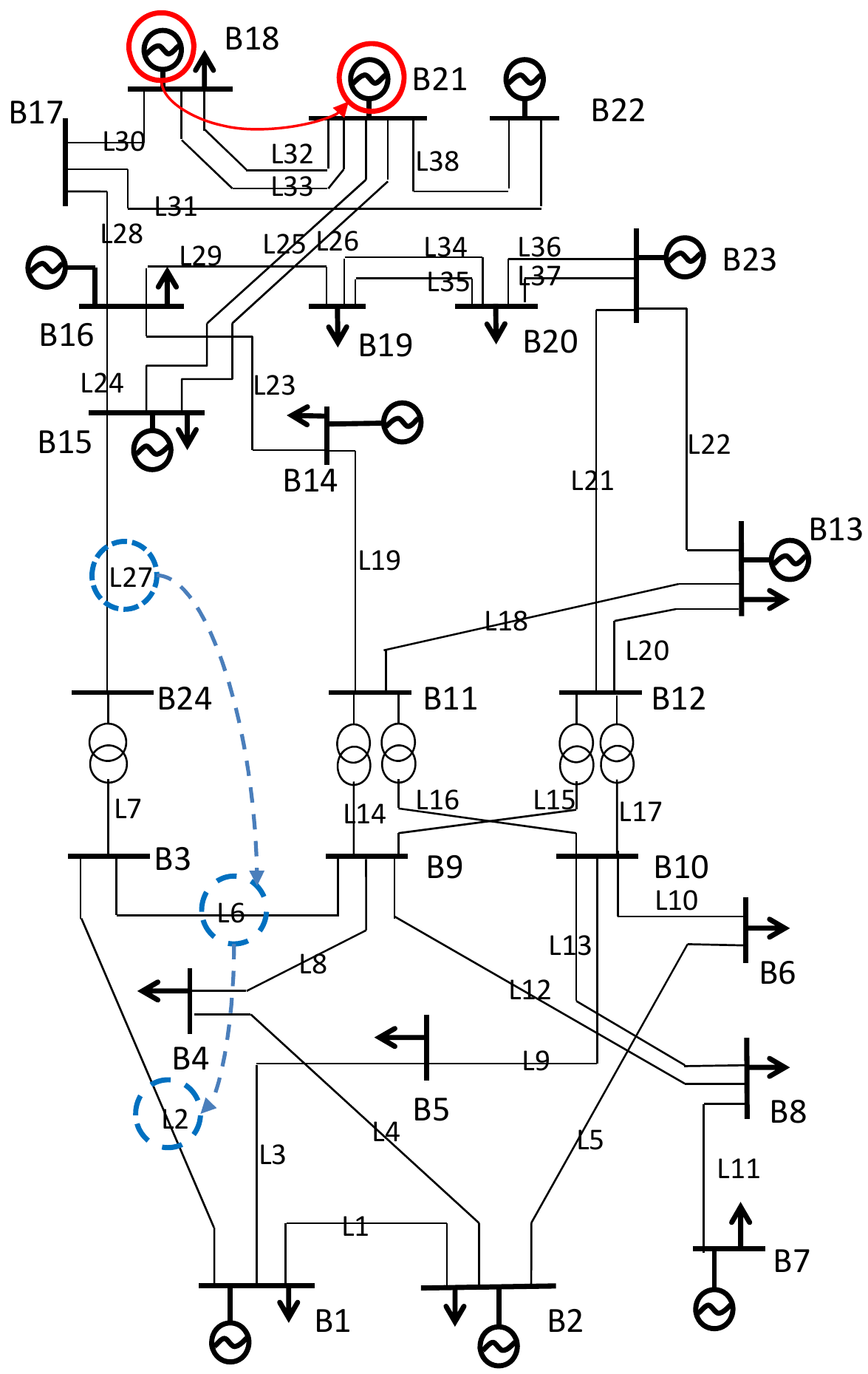}
\centerline{\footnotesize{(a)}}
		\label{fig3_1_typical}
\end{minipage}
\begin{minipage}[ht]{8.8cm}
		\centering
		\includegraphics[width=2.3in, height=3.5in]{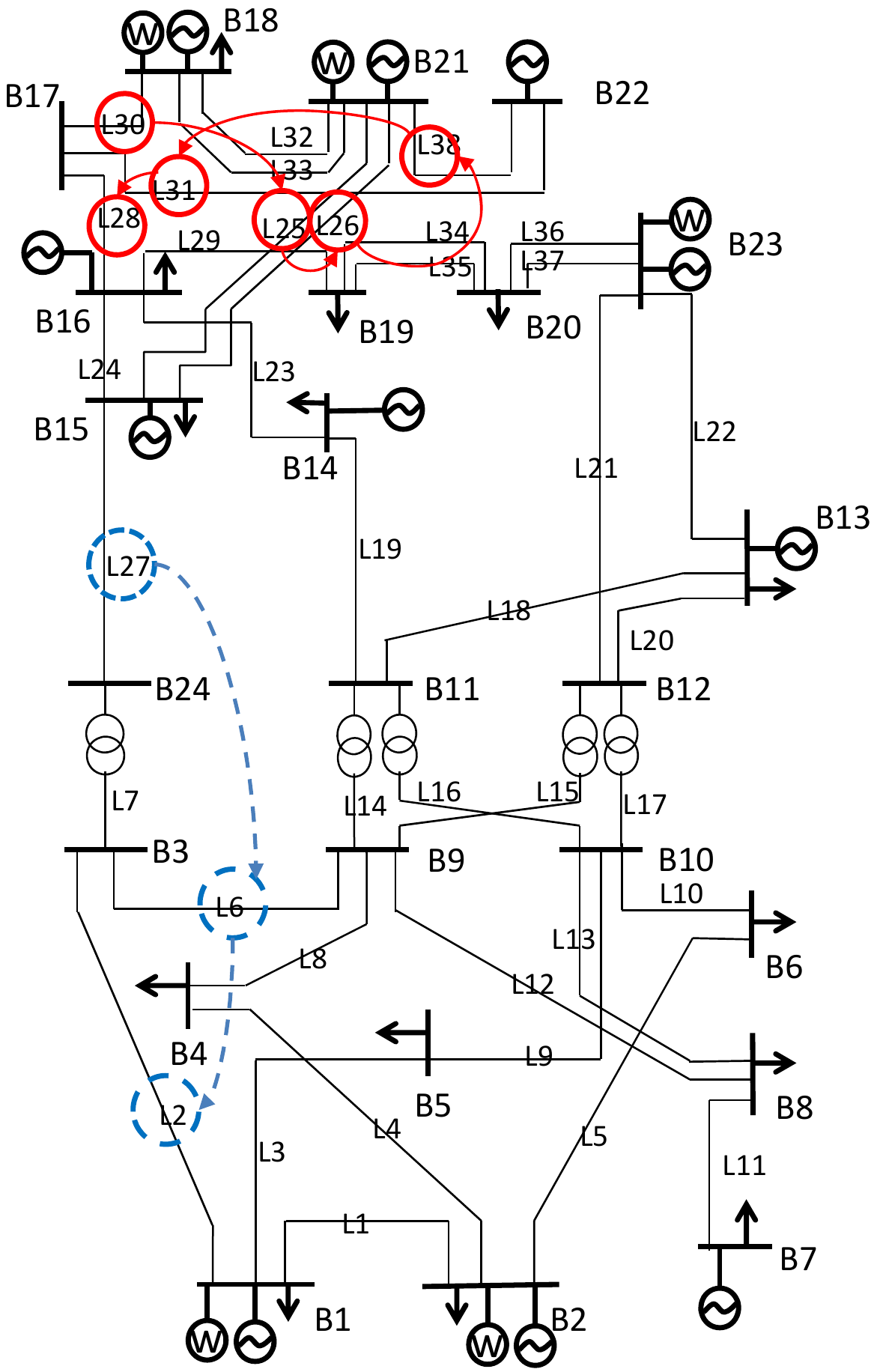}
\centerline{\footnotesize{(b)}}
		\label{fig3_2_typical}
\end{minipage}
\caption{Typical cascading paths of the IEEE RTS-79 test system. The blue dotted line represents the cascading path with the maximum expected load shedding, and the red solid line represents the cascading path with the maximum load shedding. (a) Case without wind integration. (b) Case with wind integration.}
	\label{fig3_typical}
\end{figure*}
The detailed results of some typical cascading paths are depicted in \figurename\ref{fig3_typical}.
The typical paths are selected from two aspects: the maximum expected load shedding (blue dotted line) and the maximum load shedding (red solid line).
We use `L' to represent lines and use `B' to represent buses.
From \figurename\ref{fig3_typical}, the cascading path with the maximum expected load shedding is the same with and without wind integration, which is `L27$\rightarrow$L6$\rightarrow$L2'.
This result can be explained by the fact that some cascading paths, especially some frequent cascading paths, are highly related to the topology of the system.
The cascading path with the maximum load shedding, however, is significantly different when considering wind integration.
It turns out to be a cascading path with multiple line failures: `L30$\rightarrow$L25$\rightarrow$L26$\rightarrow$L38$\rightarrow$L31$\rightarrow$L28'.
In our simulation, this cascading path appears only in the case with wind integration because the wind integration complicates the patterns of the power flow.
In some special power flow patterns, the random failure of one line will trigger a series of line failures, which will not happen without the integration of wind power.
As a result, the integration of wind generation will complicate the pattern of cascading paths.
Thus, a comprehensive analysis under massive scenarios is essential for better risk management.

From the above demonstrations, we find some new key components that could trigger severe cascading failures when considering wind power penetration (e.g. L30, L31, L38).
Such components are important to mitigate the cascading risk with with power penetration. 
Various strategies should be applied to the corresponding components, such as strengthening the branch inspection, etc.
We could also prevent the scenarios that are likely to trigger the failure of the above branches, such as changing the generations to limit the branch flow of the above branches.

\subsection{The Power System of Henan Province}
\subsubsection{System Description}
\noindent

\begin{figure*}[ht!]
	\centering
		\includegraphics[width=7.2in]{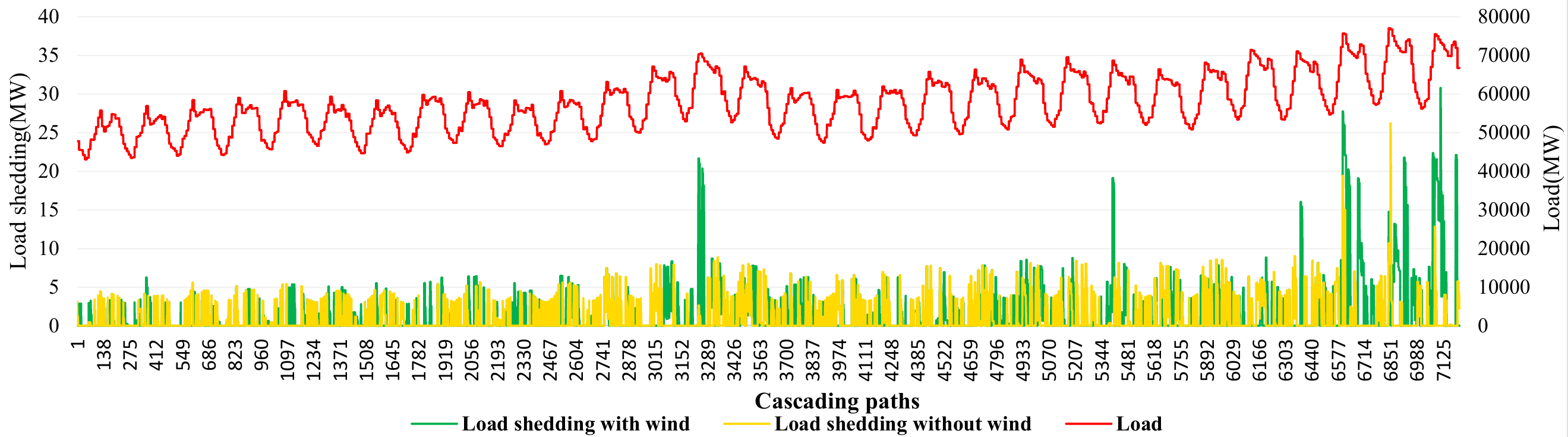}
\caption{The load shedding and the load in different cascading paths of the Henan Province power system.}
	\label{fig3_720hours}
\end{figure*}

In this work, we consider the 500 kV voltage level power system in China's Henan Province.
After some external network equivalence, we model this power system as containing 103 buses, 205 transmission lines, and 197 generators.
The cases without wind integration and with wind integration are considered.
In the case of wind integration, a total wind generation capacity of 22,557 MW  is added, which is nearly 25\% of the total generation capacity and nearly 3 times the total renewable capacity of Henan in 2018.
In addition, a total conventional generation capacity of 4,511 MW  is reduced.
The cascading failures searching of July is conducted, which is the maximum load month in Henan.
To simulate a more severe result, the actual load is enlarged by a factor of 1.05.

The minimum simulation threshold is set as $\varepsilon=10^{-10}$.
The number of cascading paths considered in each scenario is set as $m=10$.
Thus, the number of cascading paths is $744\times10=7440$.
It takes 14.31 hours to finish the simulation with the proposed method in this work.
In comparison, it takes 69.37 hours without the LSD construction technique or the fast update of the GSDF.
The proposed method saves 79.37\% computational time in this case.

\subsubsection{The Influence of Wind Power Generation}

\noindent

The load shedding results of all simulation scenarios are shown in \figurename\ref{fig3_720hours}.
The maximum load sheddings of all cascading paths increase from 26.2MW to 30.8MW when considering wind power generation.
Different from the case of the IEEE RTS-79 system, the case of Henan Province has a far lower level of load shedding with a much higher load level, even when the load is enlarged by a factor of 1.05.
This result shows the high reliability of the power system of Henan Province.
This high reliability comes from the redundancy of the installed capacity and the satisfaction of the `N-1' criterion.
Still, load shedding will happen with a small probability.
The load shedding amount is highly correlated with the total load level, without wind integration.
Regarding the high share of wind integration, both the frequency and the amount of load shedding increase significantly.
The load shedding amount is less correlated with the total load level.
The above results show that although the power system of Henan Province has a high reliability, the massive scenario cascading failure path searching is highly in need of large-scale wind power integration.

\section{Conclusions}
\label{section5_conclusions}
This paper proposes a power system cascading failure path searching approach considering wind power integration.
The proposed framework can ease the computational burden of cascading simulation under massive scenarios.
The searching framework includes the scenario generation of load and wind and the Markov chain search, where a massive long-short-long timescale simulation is conducted.
We propose an LSD-based technique to fully utilize the existing calculation results to accelerate the following calculations.
Case studies on the IEEE RTS-79 test system and empirical studies on China's Henan power system show the following:
1) The proposed LSD-based technique can significantly accelerate the simulation;
2) Wind power integration incurs more severe and more diversified cascading results and requires a complicated searching with massive scenarios.
We envision that the proposed method would serve as effective tools in power system cascading failure path searching with a high share of renewable energy integration.
\label{section6_conclusions}
\ifCLASSOPTIONcaptionsoff
\newpage
\fi



\bibliography{IEEEabrv,myReference}

%

\begin{IEEEbiography}[{\includegraphics[width=0.95in,height=1.25in,clip]{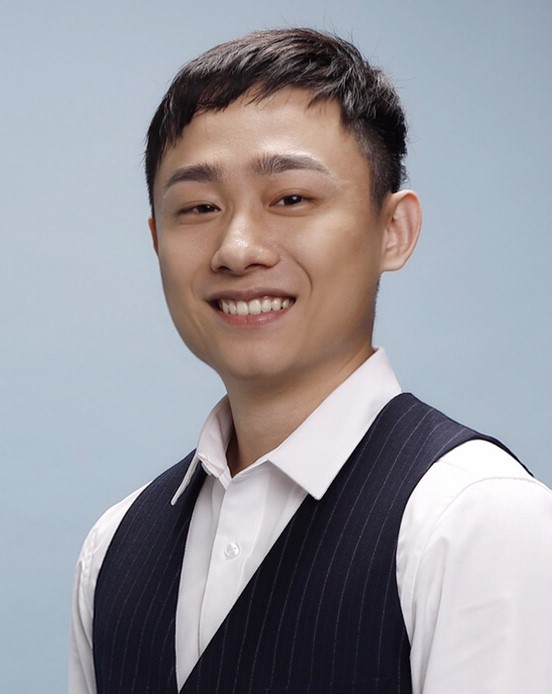}}]{Yuxiao Liu} 
(S'16) received the B.S. degree from the Electrical Engineering Department, Tsinghua University, China, in 2016.

He is currently pursuing the Ph.D. degree in Tsinghua University. 
He also holds a Visiting Graduate position with the Laboratory for Information and Decision Systems, Massachusetts Institute of Technology, Cambridge, MA, USA.
His research interests include data-driven power grid analysis and power system cascading failures modeling.
\end{IEEEbiography}
\begin{IEEEbiography}[{\includegraphics[width=0.95in,height=1.25in,clip]{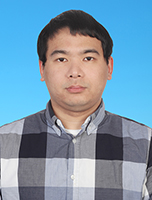}}]{Yi Wang}
(S'14-M'19) received a B.S. degree from the Department of Electrical Engineering at Huazhong University of Science and Technology (HUST), Wuhan, China, in 2014 and the Ph.D. degree at Tsinghua University, Beijing, China, in 2019. He was also a visiting student researcher at the University of Washington, Seattle, WA, USA.

He is currently a postdoctoral researcher in ETH Zurich. His research interests include data analytics in the smart grid and multiple energy systems.
\end{IEEEbiography}
\begin{IEEEbiography}[{\includegraphics[width=0.95in,height=1.25in,clip]{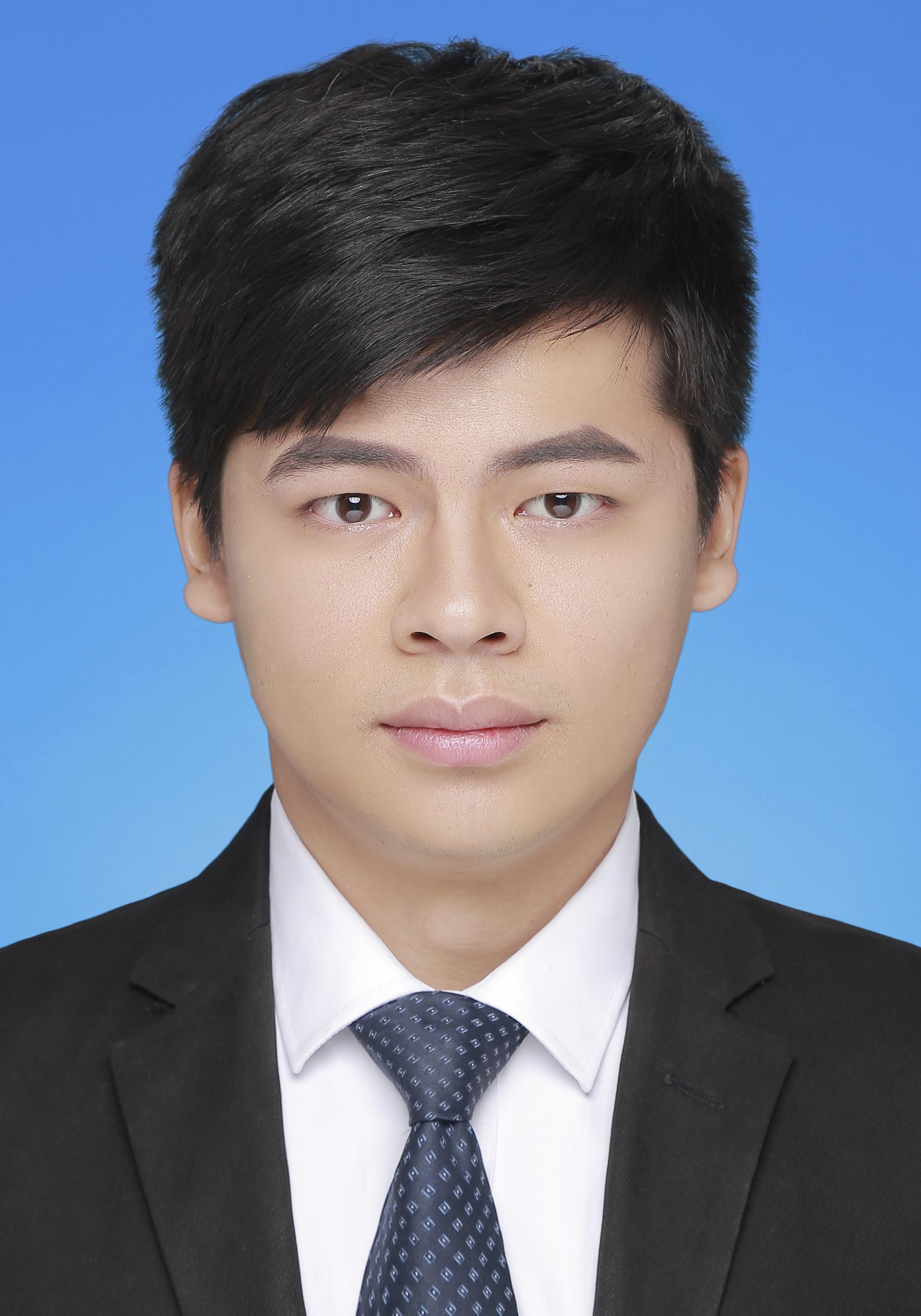}}]{Pei Yong}
(S'18) received the B.S. degree in electrical engineering in 2018 from Tsinghua University, Beijing, China, where he is currently working toward the Ph.D. degree.

His research interests include power system reliability evaluation, renewable energy and power system operation.
\end{IEEEbiography}
\begin{IEEEbiography}[{\includegraphics[width=0.95in,height=1.25in,clip]{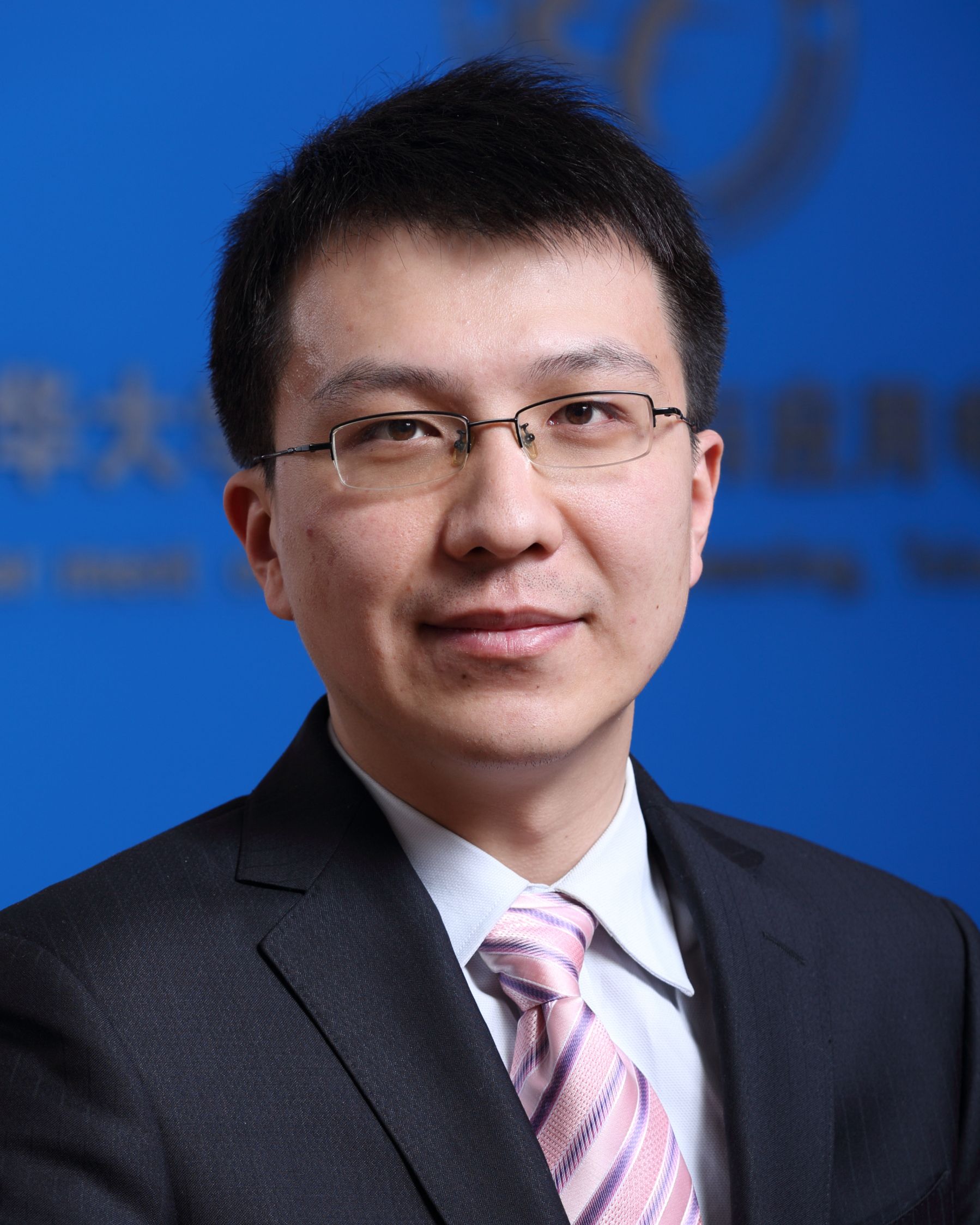}}]{Ning Zhang}
(S'10-M'12-SM'18) received both a B.S. and Ph.D. from the Electrical Engineering Department of Tsinghua University in China in 2007 and 2012, respectively. 

He is now an Associate Professor at the same university. His research interests include multiple energy systems integration, renewable energy, and power system planning and operation.
\end{IEEEbiography}
\begin{IEEEbiography}[{\includegraphics[width=0.95in,height=1.25in,clip]{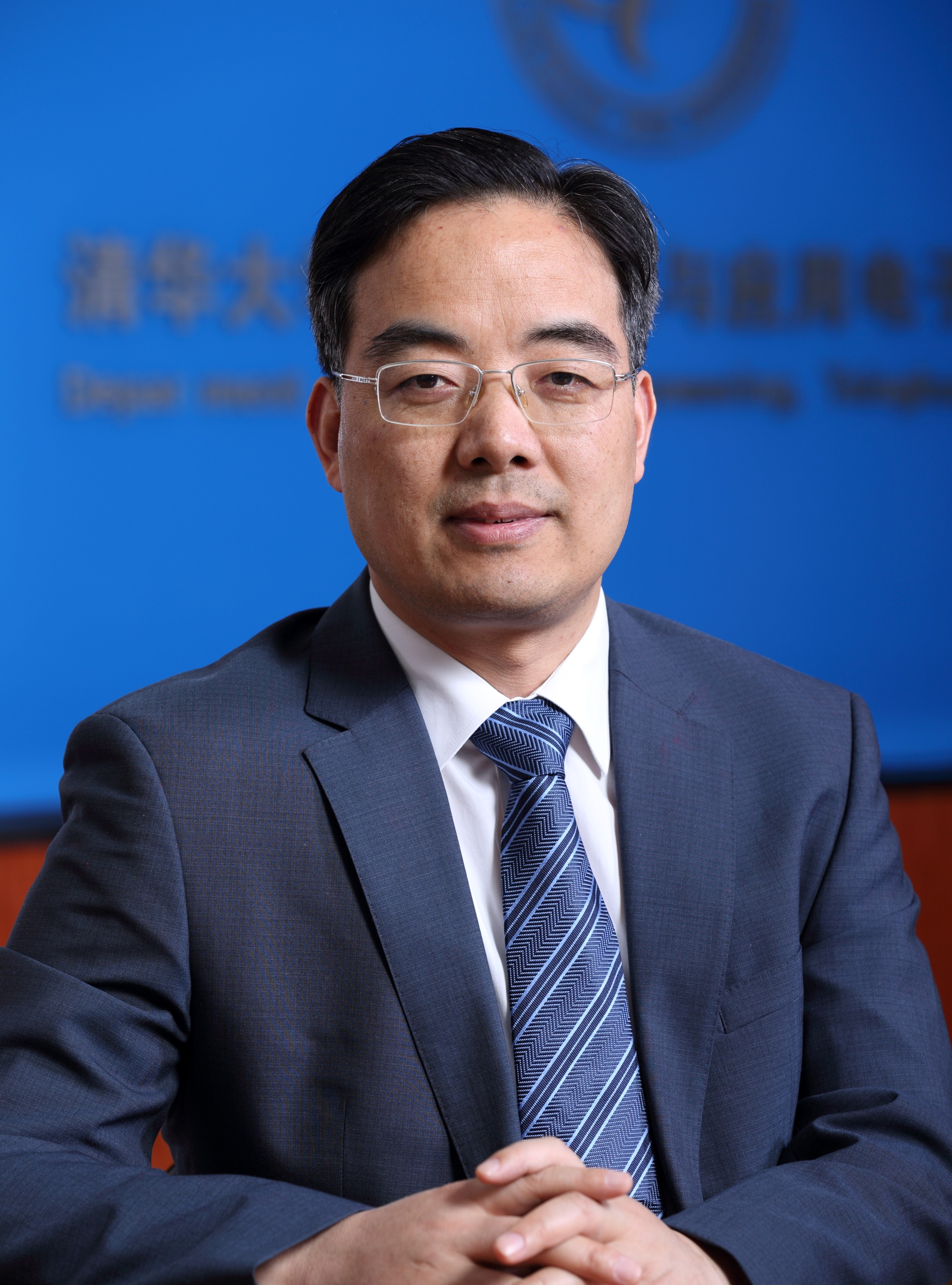}}]{Chongqing Kang}
(M'01-SM'07-F'17) received the Ph.D. degree from the Department of Electrical Engineering in Tsinghua University, Beijing, China, in 1997. 

He is currently a Professor in Tsinghua University. His research interests include power system planning, power system operation, renewable energy, low carbon electricity technology and load forecasting.
\end{IEEEbiography}
\begin{IEEEbiography}[{\includegraphics[width=0.95in,height=1.25in,clip]{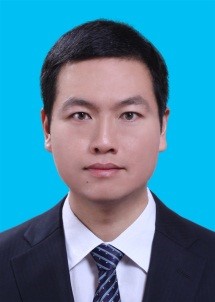}}]{Dan Lu}
 received Ph.D. degree in engineering from the Mechanical Electronic and Information Engineering Department of China University of Mining and Technology, Beijing in 2012 and 2015, respectively.

He is now working in State Grid Henan Economic Research Institute. His research interests include integrated energy system, active distribution network and renewable energy.
\end{IEEEbiography}

%
%
%




\end{document}